\documentclass[sn-basic]{sn-jnl}
\usepackage{latexsym}
\usepackage{graphicx}%
\usepackage{multirow}%
\usepackage{amsmath,amssymb,amsfonts}%
\usepackage{amsthm}%
\usepackage{mathrsfs}%
\usepackage[title]{appendix}%
\usepackage{xcolor}%
\usepackage{textcomp}%
\usepackage{manyfoot}%
\usepackage{booktabs}%
\usepackage{algorithm}%
\usepackage{algorithmicx}%
\usepackage{algpseudocode}%
\usepackage{listings}%

\def\Rbar{\mathbb{R}}

\def\Pbar{\mathbb{P}}
\newtheorem{theorem}{Theorem}
\newtheorem{proposition}{Proposition}
\newtheorem{corollary}{Corollary}

\begin{document}




\title{On the angular control of rotating lasers by means of line calculus on hyperboloids}

\author*[1]{Rudi Penne}\email{rudi.penne@uantwerp.be} 
\author[1]{Ivan De Boi}
\author[1]{Steve Vanlanduit}

\affil[1]{InViLab, Faculty of Applied Engineering, University of Antwerp,
B2020 Antwerp, Belgium}



\abstract{ 
We propose a new paradigm for modelling and calibrating laser scanners with rotation symmetry, as is the case for Lidars or for galvanometric laser systems with
one or two rotating mirrors. Instead of bothering about the intrinsic parameters of a physical model, we use the geometric
properties of the device to model it as a specific configuration of lines, which can be recovered by a line-data-driven procedure. Compared
to universal data-driven methods that train general line models, our algebraic-geometric approach only requires a few measurements.
For example, a galvanometric laser scanner with two mirrors is modelled as a grid of hyperboloids represented by a grid of
$3\times 3$ lines, providing a new type of lookup table: containing not more than 9 elements, lines rather than points, where
we replace the approximating interpolation with exact affine combinations of lines. The proposed method is validated
in a realistic virtual setting.\\
As a collateral contribution, we present a robust algorithm for fitting ruled surfaces of revolution on noisy line measurements.
}
\maketitle

\section{Introduction}
The intrinsic calibration of a sensor is typically done by determining a number of parameters in some proposed sensor model that aims to represent the physical
reality of the involved hardware \cite{cammodels,HaZi}. Often, this strategy implies non-flexible models with unstable parameter values
(\cite{sensitivitycalspheres,sensitivityLai}, Chapter 3 in \cite{phd_dvhamme}). In spite of its rich tradition and literature, 
calibration remains a tedious and time consuming task, to be repeated when conditions change,
not always obtaining the required accuracy.
The shortcomings of the calibration by matching a rigid physical device model have recently been admitted by leading scientists in the field \cite{multiparams}.
The inaccuracies and instabilities inherent to the current calibration procedures are troublesome in
applications where intrinsic localization, registration and sensor fusion are involved \cite{RadarCamFusion}.
And last but not least,  intrinsic calibration procedures based on a physical model cope with the determination of physical parameters that can rarely be measured directly, and
are moreover rather virtual than physical, due to the idealised abstract nature of the model.

An alternative strategy is the so-called universal-model-based method, 
which considers a sensor as a black box that connects
its control variables (camera pixel coordinates, mirror angles for laser reflection,\ldots ) to the observed
world. The calibration of this mapping is established by a data-driven procedure, requiring
the availability of sufficiently large datasets that allow interpolation \cite{multiparams,interpollinefunc}, or
lookup tables \cite{LUTcalib},
or the training of neural networks or Gaussian processes \cite{NewSensors,DDGalvano1,DDHyperspectral,WrappedGaussian,IvanConstrGP}.
An important issue of this approach is that it requires world point clouds with reliable coordinates, which
significantly cover the work space.

In this article we make use of geometric sensor models, assigning a world line 
for each sensor query
\cite{raxel,raygeo,ponce2,LearningAV}. Lines naturally present the way how many sensors observe the world (beams of light). 
Our approach still assumes a specific model, given by a {\em line variety\/} (in the
algebraic geometric sense). But it bypasses the intrinsic physics of the device. The lines that belong to this line model, can be obtained by direct measurements, as opposed to the parameters of a physical model-based calibration. A practical drawback of line measurements might be that they require to determine the position of several collinear points. 
However, this extra work is awarded with the possibility
to reduce noise and outliers for point measurements by means of robust line fitting 
\cite{FB}.
Furthermore, a line model provides stable transformations to other reference frames (extrinsic
calibration \cite{ExtrLaser,NPnP,generalpose}). 
The  calibration algorithms as presented in
\cite{posenoncentral,ponce1,ponce2} are line-model-based, but they 
still use (non-obvious) parameter models
and appear to be too complicated for practical purposes. Alternatively, some authors avoid restrictions
on the involved line variety, calibrating a universal line model through a data-driven learning process
\cite{DDGalvano2,IvanSemiDriven,LearningAV}. These universal approaches have the advantage that they are not based on geometric assumptions (except for the straight line assumption),
but they need the availability
of a large set of line measurements and suffer from a lack of theoretical accuracy guarantees. 
This paper demonstrates the profit of proposing a specific type of line variety as sensor model, supported by natural geometric assumptions. In this way, we compromise between model-line-based 
and data-line-based approaches. 

In many applications we use a sensor that corresponds to a two-dimensional line variety (a camera with two pixel coordinates, 
a laser scanner with two control parameters,\ldots), which is called a {\em line congruence}
\cite{twoslit,PoStu,LineCongr}.  
 
In this article we elaborate line-model sensors with rotational symmetry, as it is the case for scanners with rotating lasers (Lidar)
or for a galvanometric laser scanning system where
a fixed laser beam is reflected by one or two rotating mirrors. We prove that the corresponding line varieties are covered by ruled 
quadratic surfaces of revolution
As an important application,
we present a novel, fast and efficient line-based calibration procedure for a
two-mirror galvanometric laser scanner (2M-GLS) (Figure~\ref{fig_vibro}). These laser scanners appear in several applications \cite{PosGalvano,twomir1,twomir2,twomir3}
due to their ``good characteristics of high deflection speed, high positioning repeatability and concise structure'' \cite{DDGalvano2}. 
For the majority of the publications, the authors
restrict to situations where a 2M-GLS measures a plane or a two-dimensional surface \cite{PlaneGalvano}. In such situations there is no need to go beyond point-based calibrations. However, for a complete 3D-range, sensor calibration
must provide the 3D-line (in some reference frame) for each selected pair of rotation angles of the
two mirrors that reflect a fixed incoming laser beam.
Model-based methods for the 3D-calibration of a 2M-GLS are given by 
\cite{modelGallaser,modelGalvano}. However, these methods have to determine (too) many
parameters of (a model of) the device geometry. They cope with the disadvantages
that are listed at the beginning of the introduction (unstable and tedious to implement),
giving rise to
non-convex optimization problems that suffer from local minima.
In \cite{DDGalvano2,IvanSemiDriven} the authors
propose to calibrate a 2M-GLS by a data set of line
measurements, which is more related to our approach. However, their method completely differs from
the proposed procedure, because they calibrate a universal line model through a statistical learning process, 
without bothering about the algebraic and geometric structure of the involved
line congruence. 

In order to present the mathematical tools for this article in a self-contained manner,  
we provide the complete description of the hyperboloids or cones of revolution that are obtained by the laser reflections in the case of one rotating mirror (Section~\ref{sec_refl}). This leads to the specific
Pl\"ucker coordinates of these laser reflections as presented in Section~\ref{sec_Pl}. As a collateral
application, we present a robust algorithm in the Appendix for recovering a ruled surface of revolution from noisy line data. 
An important contribution and innovation in this article
appears in Section~\ref{sec_affpl}, where we derive a representation of the lines of one hyperboloid of revolution as a stable one-parameter combination of three
generating lines, directly related to the angular variable that controls the mirror rotation.
This result was accomplished thanks to the rational parameterization of affine combinations on
the circle as presented in Section~\ref{sec_affcirc}. 
In Section~\ref{sec_algo} we show how this result yields an algorithm to predict laser reflections, first for one rotating mirror, and then extended to the
concept of a three by three hyperboloid grid for modelling and calibrating
a two-mirror galvanometric laser scanner (Section~\ref{sec_vibro}).

We believe that this article offers a novel and fundamental contribution to the field of sensor modelling and calibration, especially useful for laser scanners with rotational components. 
We show how certain sensors can be represented by line congruences that on their turn can be represented by a limited  base set of lines. For example, a galvanometric laser scanner with one mirror can be represented by 3 lines, and in the case of two mirrors by 9 lines. We discovered how to generate the whole line congruence from these bases by a linear line calculus that is directly related to the angular control parameters. The correctness of our 
{\em hyperboloid grid method\/} is validated by mathematical proofs,
the accuracy and stability by the synthetic experiments in Section~\ref{sec_exp}, by means of a
virtual 2M-GLS that simulates real world hardware. We observe a very
accurate and precise performance, as well as a favourable comparison with the data-based calibration
of \cite{IvanConstrGP} that is known to outperform physical parameter models and to match
other statistical learning models.
This success is mainly explained by the stability of the calculus on hyperboloid grids introduced in Section~\ref{sec_affpl} (validated in Section~\ref{sec_exp}),
due to the use of a stable rational parameterization to represent the mirror rotation. 
In addition, the line-based algorithm for fitting ruled quadrics of revolution, as presented in the Appendix, definitely improves the robustness of the proposed calibration method.

\section{Line reflections by rotating mirrors}

\label{sec_refl}

This section describes the well known geometry of the reflections of a fixed incoming laser
beam with a mirror that rotates about a fixed axis, offering the opportunity to introduce our terminology. We assume that this rotation axis $A$,
the laser beam $L$ and its reflections can be modelled by (straight) spatial lines, and the mirror by a (flat) plane that contains the axis $A$.
Typically, only one side of this rotating 
performs mirror reflection, such that it makes no sense to allow a rotation angle range that exceeds $180^\circ$. For most physical devices, this range is even more restricted. 

A particular position of the rotating mirror ${\cal M}(n)$ is determined by its unit normal $n$, which is supposed to point in the sense of mirror reflection. So, if the laser $L$ is directed
by the unit vector $r_L$, compatible with the incoming orientation, 
then the reflected line $R(n)$ has direction vector $r_n$ (according
to the orientation of reflection): 
\begin{equation}r_n = r_L - 2 (r_L\cdot n) n.\end{equation}

\begin{figure}[htbp]
\begin{center}
\includegraphics[width=6cm]{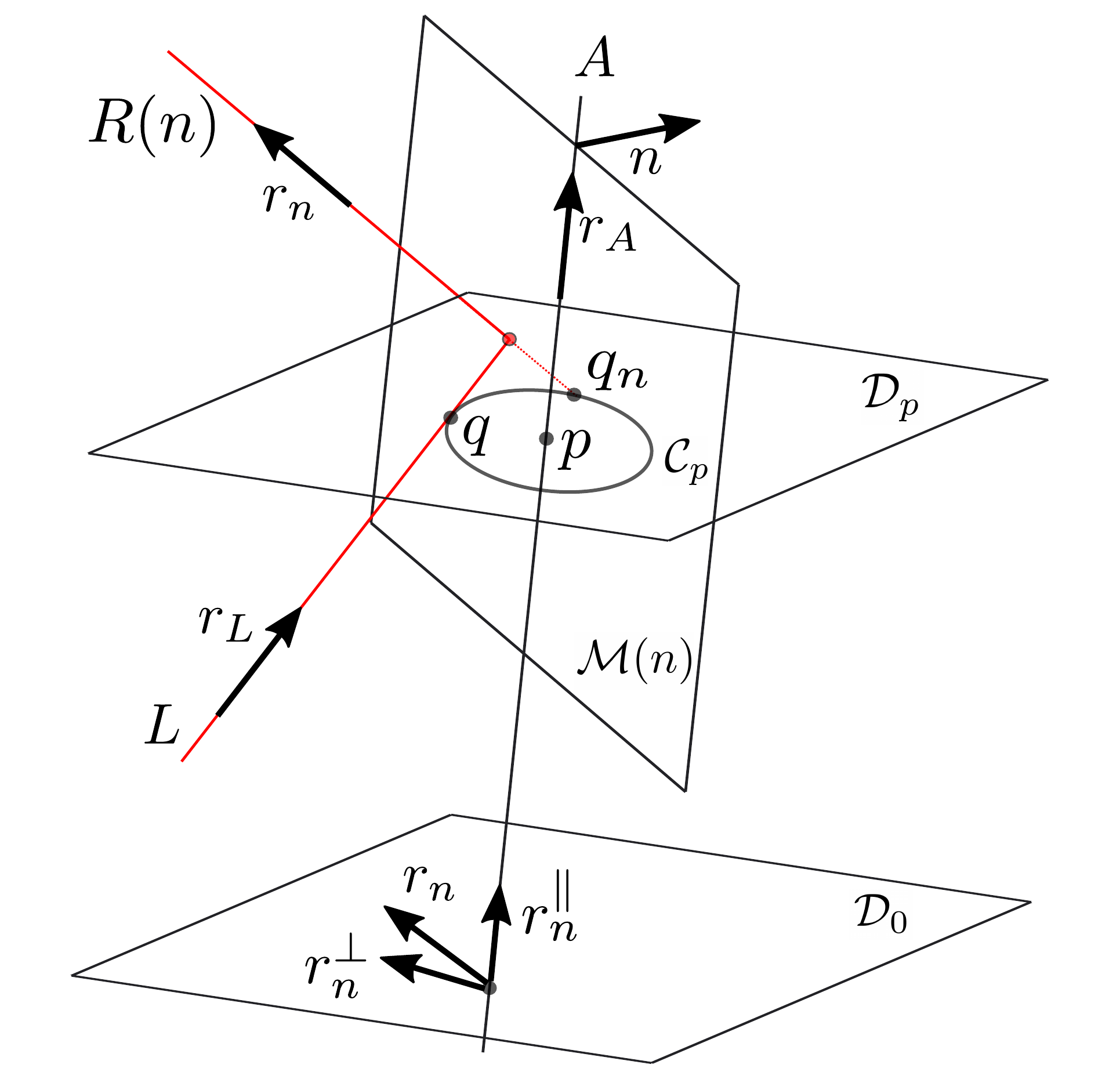}
\includegraphics[width=5cm]{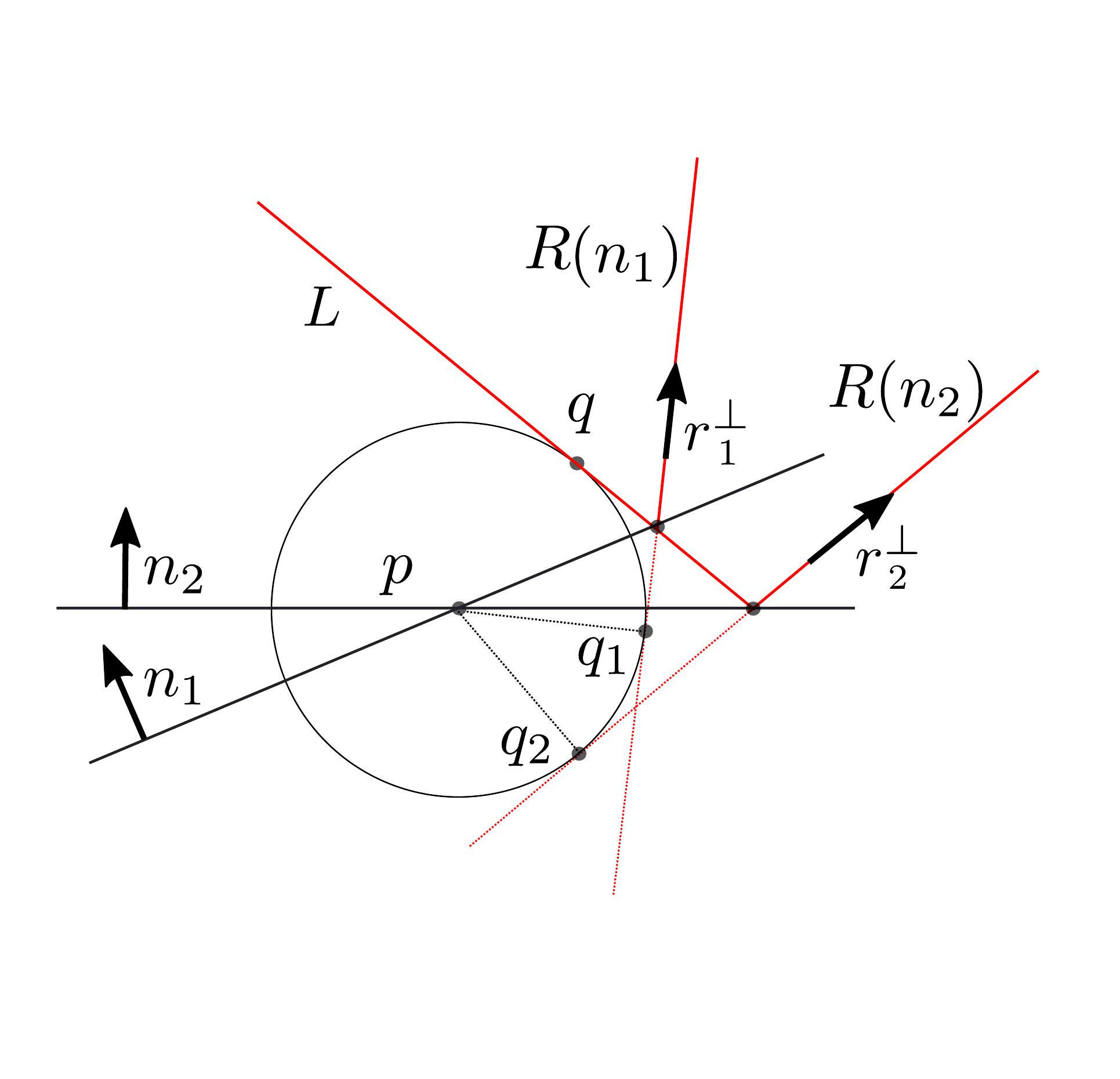}
\end{center}
\caption{Both the incoming as the reflected laser are rulers of the
same hyperboloid of revolution.}
\label{fig_vars}
\end{figure}

Further, let us agree that the normalised direction of the mirror rotation axis $A$ is denoted by $r_A$, and the plane through the origin and perpendicular to $A$ by ${\cal D}_0$
(Figure~\ref{fig_vars}). Notice that ${\cal D}_0$ differs from the plane containing
$r_L$ and $r_n$, unless the incoming laser happens to be orthogonal to $A$. For this reason we decompose (incident and reflected) line directions $r$ in
a component along $A$ and a component perpendicular to $A$ (in ${\cal D}_0$):
\begin{equation}r = (r\cdot r_A)r_A + r^\perp = r^\parallel + r^\perp .\end{equation}
We will always assume that the incident laser hits (is not parallel to) the mirror, so $r^\perp$ is not the zero vector.
Let $R(n_1)$ and $R(n_2)$ be reflections of the same incident laser $L$ for different mirror positions ${\cal M}(n_1)$ and ${\cal M}(n_2)$ during the rotation about axis $A$. Let
$r_1$ and $r_2$ abbreviate $r(n_1)$ and $r(n_2)$ respectively (Figure~\ref{fig_vars}).

\begin{proposition}
\label{prop_georefl}
${}$
\begin{enumerate}
\item $r_1^\parallel = r_2^\parallel$. 
\item $R(n_1)$, $R(n_2)$ and $L$ cross $A$ at equal distance, sharing a common closest point
$p$ on $A$. So, if  ${\cal D}_p$ denotes the plane through $p$ and
perpendicular to $A$, then $p$ is the center of a circle ${\cal C}_p$ in ${\cal D}_p$, intersecting
$L$, $R(n_1)$ and $R(n_2)$ in $q$, $q_1$ and $q_2$, respectively. Furthermore:
\begin{equation}
\left<{q_1-p,q_2-p}\right>
=\left<{r_1^\perp, r_2^\perp}\right>= 2\left<{n_1, n_2}\right>.
\label{eq_double}
\end{equation}
\end{enumerate}
\end{proposition}

Proposition~\ref{prop_georefl} implies that all line reflections of a fixed 
laser by a continuously rotating mirror (over some angle range) can be equally well obtained by the continuous rotation of the first reflected line (over the double range). 
It is a well known geometric fact that
the rotation of a line around a given fixed axis $A$ sweeps a one-sheeted hyperboloid of revolution ${\cal H}$\cite{quadrics}. ${\cal H}$ can be considered as a union of lines but equally well as the union of circles (perpendicular to $A$). The smallest of these circles, ${\cal C}_p$ in Proposition~\ref{prop_georefl}, is called the {\em gorge circle\/} of this surface of revolution. We conclude in the following theorem, where we take care for the singular situations:

\begin{theorem}
\label{th_hyperb}
If the incoming laser beam $L$ is not perpendicular to the mirror rotation axis $A$,
and if $L\cap A = \emptyset$ then the reflected lines
belong to one system of rulers of a one-sheeted hyperboloid of revolution, ${\cal H}(L,A)$, 
completely
determined by $L$ and $A$. Indeed, the gorge circle of ${\cal H}(L,A)$ is given by ${\cal C}_p$, and its {\em pitch\/} $\rho$ by:
$$\rho = r_n\cdot r_A = -r_L\cdot r_A.$$
The incoming laser $L$ belongs to the second system of rulers on ${\cal H}(L,A)$. If $L$ intersects $A$, then ${\cal H}(L,A)$ degenerates into a cone, or even into a flat pencil if $L$ happens to intersect $A$
perpendicularly. Finally, if $L\perp A$ and $L\cap A =\emptyset$, then ${\cal H}(L,A)$ degenerates into the set of tangents to ${\cal C}_p$ in ${\cal D}_p$. 
\end{theorem}

\begin{figure}[htbp]
\begin{center}
\includegraphics[width=7cm]{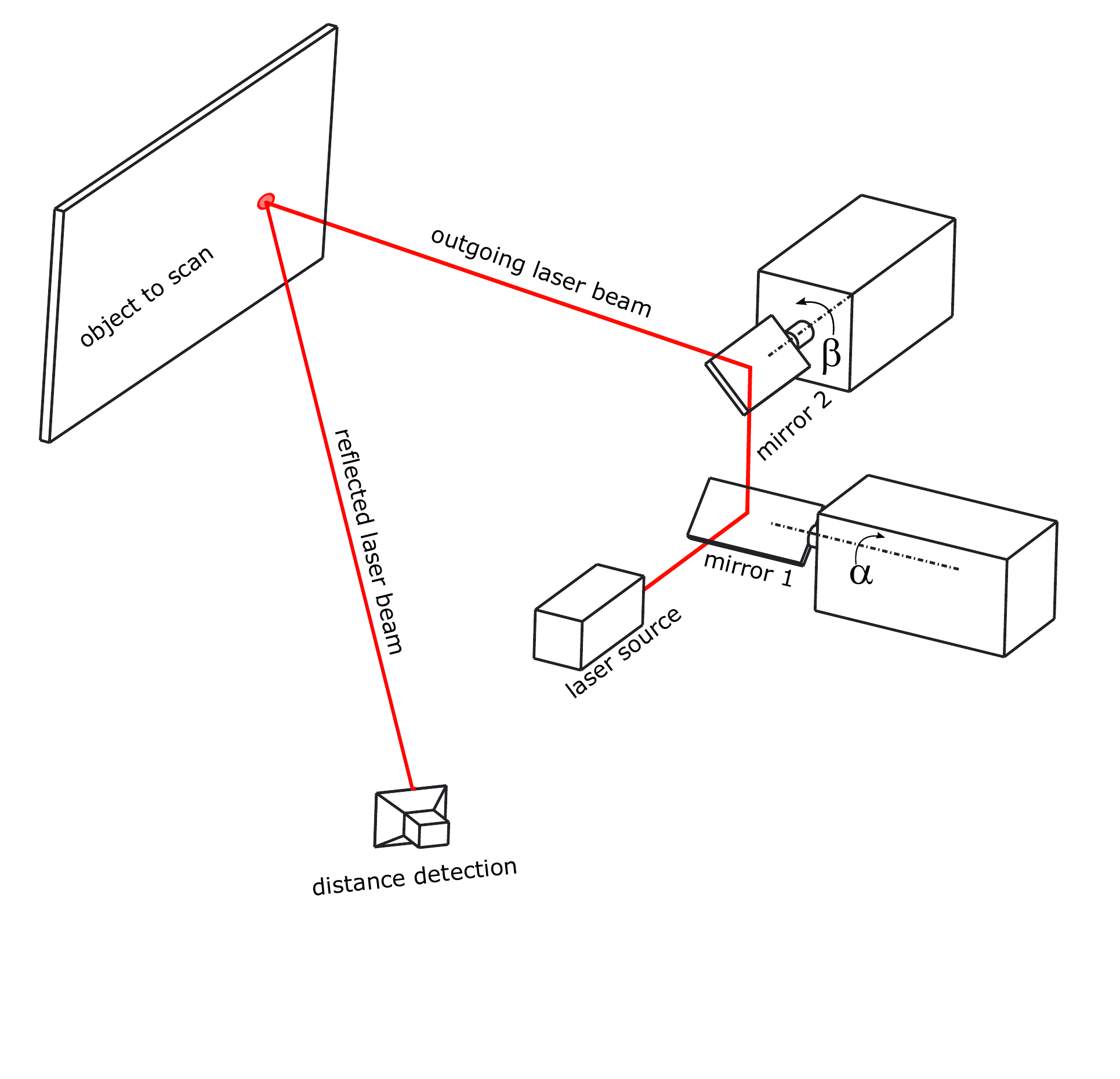}
\end{center}
\caption{The setup of a two-mirror galvanometric laser scanner (2M-GLS).}
\label{fig_vibro}
\end{figure}

The next step is to consider a 2M-GLS, a sensor consisting of a single fixed laser $L$ that is internally reflected by two sequential mirrors, 
each rotating about an individual axis, denoted by $A$ and $B$ in order of reflection. The control
of the individual rotating mirrors is typically galvano-driven, allowing two independent user parameters, denoted by
$\alpha$ and $\beta$ respectively (Figure~\ref{fig_vibro}).

Note that we only observe the outgoing lasers of the galvanometer after the second reflection by
the mirror ${\cal M}(B,\beta)$ that rotates about the axis $B$. An arbitrary value of
the parameter $\alpha$ that controls the position of the first mirror ${\cal M}(A,\alpha)$, generates
a reflection $L(\alpha)$ of the initial laser $L$, which is on its turn the incident laser for the
rotating mirror ${\cal M}(B,\beta)$. Because the laser line that is generated by a 2M-GLS is
determined by a pair of angle settings $(\alpha,\beta)$, it can be denoted by $R(\alpha,\beta)$.
Theorem~\ref{th_hyperb} translates into:

\begin{theorem}
\label{prop_hypernet}
The outgoing lasers $R(\alpha,\beta)$ of a 2M-GLS lie on a family of (possibly degenerate) co-axial hyperboloids of revolution
${\cal H}(L(\alpha),B)$, each of which is generated by an individual laser $L(\alpha)$ that is reflected by rotating the second mirror ${\cal M}(B,\beta)$.
\end{theorem}

Varieties of lines with two degrees of freedom, such as the line reflections produced by
two rotating mirrors, are called {\em line congruences\/} \cite{LineCongr}. In our case
we coin the name {\em two-mirror congruence\/}.

\bigskip
\noindent
{\bf Warning:} The centres of the different hyperboloids ${\cal H}(L(\alpha),B)$, being the points
$p(\alpha)$ on $B$ with minimal distance to $L(\alpha)$, are not equal (except in degenerate
cases). Therefore, the congruence of laser lines emitted by a 2M-GLS does not constitute
a {\em linear line congruence\/} \cite{PW,PoPeRa,PoStu,LineCongr}. 

If we consider the intermediate state of the sensor, after the first rotating mirror 
${\cal M}(A,\alpha)$, 
then the reflected beams of the incoming laser $L$ also lie on
a hyperboloid, ${\cal H}(L,A)$. If we fix the second mirror at position $\beta_1$, 
${\cal M}(B,\beta_1)$, while
rotating the first mirror, then we observe the sensor emitting a mirror reflection of ${\cal H}(L,A)$ by
${\cal M}(B,\beta_1)$. Of course, this mirror image is also a one-sheeted hyperboloid of revolution, denoted
by ${\cal H}(L,A,B,\beta_1)$, containing the doubly reflected laser beams $R(\alpha,\beta_1)$ (with
varying $\alpha$). Consequently, we can be more specific about the description of the
two-mirror congruence as given by Theorem~\ref{prop_hypernet}.

\begin{corollary}
The outgoing lasers $R(\alpha,\beta)$ of a 2M-GLS belong to a congruence that can be considered
as the disjoint union of either of the following two systems of hyperboloids of revolution:
\begin{enumerate}
\item A system of (co-axial) hyperboloids, each of them determined
by lines $R(\alpha_1,\beta)$ with constant $\alpha_1$.
\item A system of hyperboloids with each of them determined by lines $R(\alpha,\beta_1)$
with constant $\beta_1$.
\end{enumerate} 
\label{cor_hypernet}
\end{corollary}

Observe that the axes of the second system of hyperboloids in Corollary~\ref{cor_hypernet}
sweep an additional hyperboloid of revolution, not participating in the two-mirror congruence,
but sharing its axis with the hyperboloids of the first system. 

\section{Pl\"ucker coordinates of reflections of a single laser by a rotating mirror}
\label{sec_Pl}

We refer to  \cite{PW} for an introduction to line coordinates and line geometry
in a projective geometric setting, or to \cite{quadrics} for a Euclidean definition of line
coordinates. In our context it is natural to work over
the real numbers $\Rbar$ as a base field.
A line $R$ in Euclidean 3-space is determined by its direction vector $r$ and a point $q$. In order to
get rid of the randomness in selecting $q$ on $R$, we replace $q$ by the moment $m=q\times r$, 
which is independent of the choice of $q$ on $R$, and only depends on the scale of $r$. Observe
that $q\times k r = k (q\times r)$, so the sixtuple $(r,m)$ gives well defined homogeneous
coordinates for $R$, {\em Pl\"ucker coordinates\/},
mapping this line in 3-space to a point $\pi(R)$ in $\Pbar^5$. Furthermore,
since $r\cdot m =0$, this point belongs to the so-called {\em Klein quadric\/} ${\cal K}$ in
$\Pbar^5$:
$${\cal K} = \{(x_1:x_2:x_3:x_4:x_5:x_6)\in\Pbar^5\, |\, x_1x_4+x_2x_5+x_3x_6=0\}.$$
It can be proven that every point of ${\cal K}$ either represents the Pl\"ucker coordinates of
a Euclidean line, or it represents a ``line at infinity'' (where $x_1=x_2=x_3=0$).\\
Finally, for a Euclidean line $R$, we can tie down the random homogeneous factor by normalizing
its direction vector: $||r||=1$. To avoid the final ambiguity, we will always assume that each
line $R$ has a given orientation.

\bigskip
A major objective of this paper is to control the Pl\"ucker coordinates of the laser reflections by
means of the rotation angle of the mirror. In order to present the algebraic calculus of laser
reflections more easily, we will assume for the moment that the origin coincides with
the point $p\in A$ that has minimal distance to the incoming laser $L$, implying that
${\cal D}_0={\cal D}_p$ 
(Section~\ref{sec_refl}). Later we will see that this choice does not affect the derived formulas.

Recall form Proposition~\ref{prop_georefl} that the laser reflections $R(n)$ 
corresponding to different positions ${\cal M}(n)$ of the rotating
mirror share an identical pitch $\rho=r_n\cdot r_A$, where we assume that the direction vectors
$r_n$ (of $R(n)$) and $r_A$ (of the rotation axis $A$) are normalised and oriented such that
$\rho>0$. Consequently, the projections $r^\perp$ on ${\cal D}_0$ of the reflected directions $r$
all have identical norm $||r^\perp||=\sqrt{1-\rho^2}$. Furthermore, Proposition~\ref{prop_georefl} implies that each $r^\perp$ is perpendicular to $q_n-p = q_n = R(n)\cap {\cal D}_0$ ($=$
closest point of $R(n)$ to the axis $A$). If $L$ does not intersect $A$, all these points $q_n$ belong to
the gorge circle ${\cal C}_0$ of the hyperboloid ${\cal H}(L,A)$ with radius $\sigma_0=||q||$
(Figure~\ref{fig_vars}). Finally, recall that the relative (oriented) angles 
of rotation between the reflected lines $R(n)$ are determined by the mirror rotation ${\cal M}(n)$:
$$\left<{r_1^\perp, r_2^\perp}\right> =
\left<{q_1,q_2}\right>= 2\left<{n_1, n_2}\right>.$$
Our next observation is that the moments $m_n=q_n\times r_n$ of the reflected lines $R(n)$ appear
to behave in a similar way as the directions. Except for the special case where $L$ intersects $A$ 
(in $p=q_n=$the origin), implying that $m_n$ is the zerovector.

\begin{proposition}
\label{prop_mom}
Assume the previous notations and assumptions, in particular the origin is located
at $p\in A$, and assume that $L$ does not intersect $A$. 
Then the laser reflections $R(n)$ corresponding to different positions ${\cal M}(n)$ of the rotating
mirror share an identical {\em moment pitch\/} $\mu=m_n\cdot r_A$.
Furthermore, if $m_n^\perp = m_n - \mu r_A$ denotes the moment projection on ${\cal D}_0$, then
$m_n^\perp$ is parallel to $r_n^\perp$ with $||m_n^\perp||=\sigma_0\cdot\rho$ (where $\sigma_0=||q_n||=||q||$ is the radius of ${\cal C}_0$).
\end{proposition}

\noindent{\bf Proof.}
Due to our choice of the origin, $q$ belongs to ${\cal D}_0$, where it is orthogonal
to $r_n^\perp$. Recall that $R(n)$ is oriented by $r_n$ conform to the sense of the reflection, and that the mirror axis $A$ is orientated by $r_A$ such that $\rho = r_n\cdot r_A=-r_L\cdot r_A > 0$.
Notice that in case the skew oriented lines $A$ and $R(n)$ cross ``positively'', which means that
the undercrossing line passes the overcrossing from left to right, $r_n^\perp$ is obtained by a clockwise quarter turn from in $q_n$ in ${\cal D}_0$ as viewed from $r_A$. Also note that for each
mirror position ${\cal M}(n)$ the crossing sign of $R(n)$ relative to $A$ is the same, namely the opposite of
the crossing sign of $L$ and $A$.  

So, due to the right-hand-rule for the orientation of the cross product $m_n=q_n\times r_n$, and
because $r_n\cdot r_A >0$, we see that $\mu=m_n\cdot r_A <0$ if and only if $R(n)$ crosses
$A$ positively. 
Because both $r_n\perp q_n$ and
$m_n\perp q_n$ the projections $m_n^\perp$ and $r_n^\perp$ are aligned in
${\cal D}_0$. So, $m_n^\perp = k r_n^\perp$, where $k>0$ if and only if 
$\mu <0$.
We conclude that the sign of the moment pitch is the same for every laser reflection $R(n)$.

Let us now compute the size of the moment pitch:
\begin{eqnarray*}
\mu &=& (q_n\times r_n)\cdot r_A\\
&=& (q_n\times (r_n^\perp + \rho r_A))\cdot r_A\\
&=& (q_n\times r_n^\perp)\cdot r_A\\
&=& \pm ||q_n\times r_n^\perp||
\end{eqnarray*}
where we used that $(q_n\times r_A) \perp r_A$ and $(q_n\times r_n^\perp)\parallel r_A$.
But $q_n\perp r_n^\perp$, so 
$$|\mu| =||q_n||\cdot ||r_n^\perp||=||q_n||\sqrt{1-\rho^2},$$
which finishes the proof that $\mu$ is independent from the mirror position.

In addition, $||m||=||q_n||\cdot||r_n||=||q_n||=||q||$, whence 
$$||m_n^\perp||^2=||m_n^2||^2 - |\mu|^2 = ||q||^2 \rho^2.$$
\hfill $\blacksquare$

\begin{figure}[htbp]
\begin{center}
\includegraphics[width=7cm]{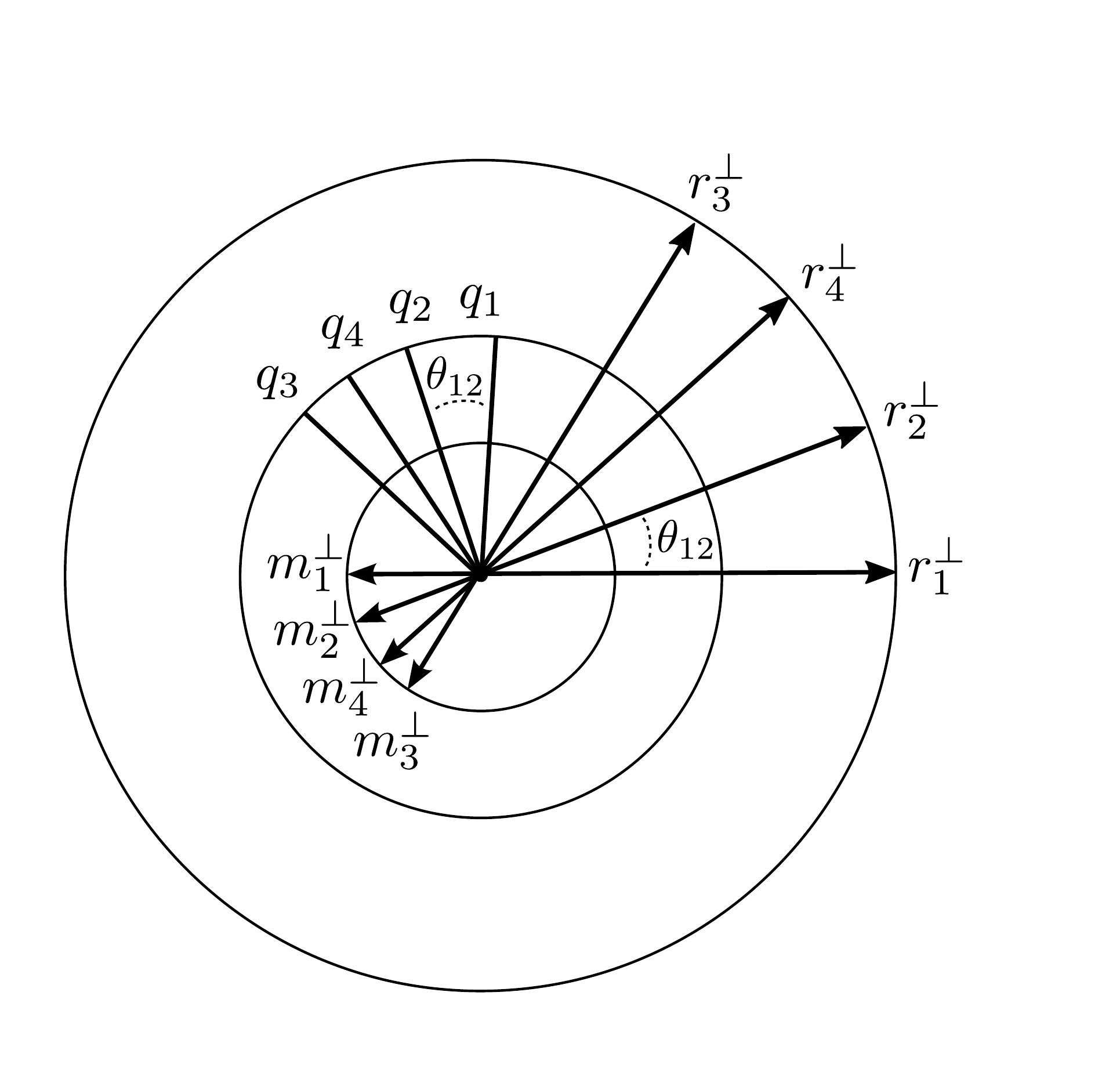}
\includegraphics[width=4cm]{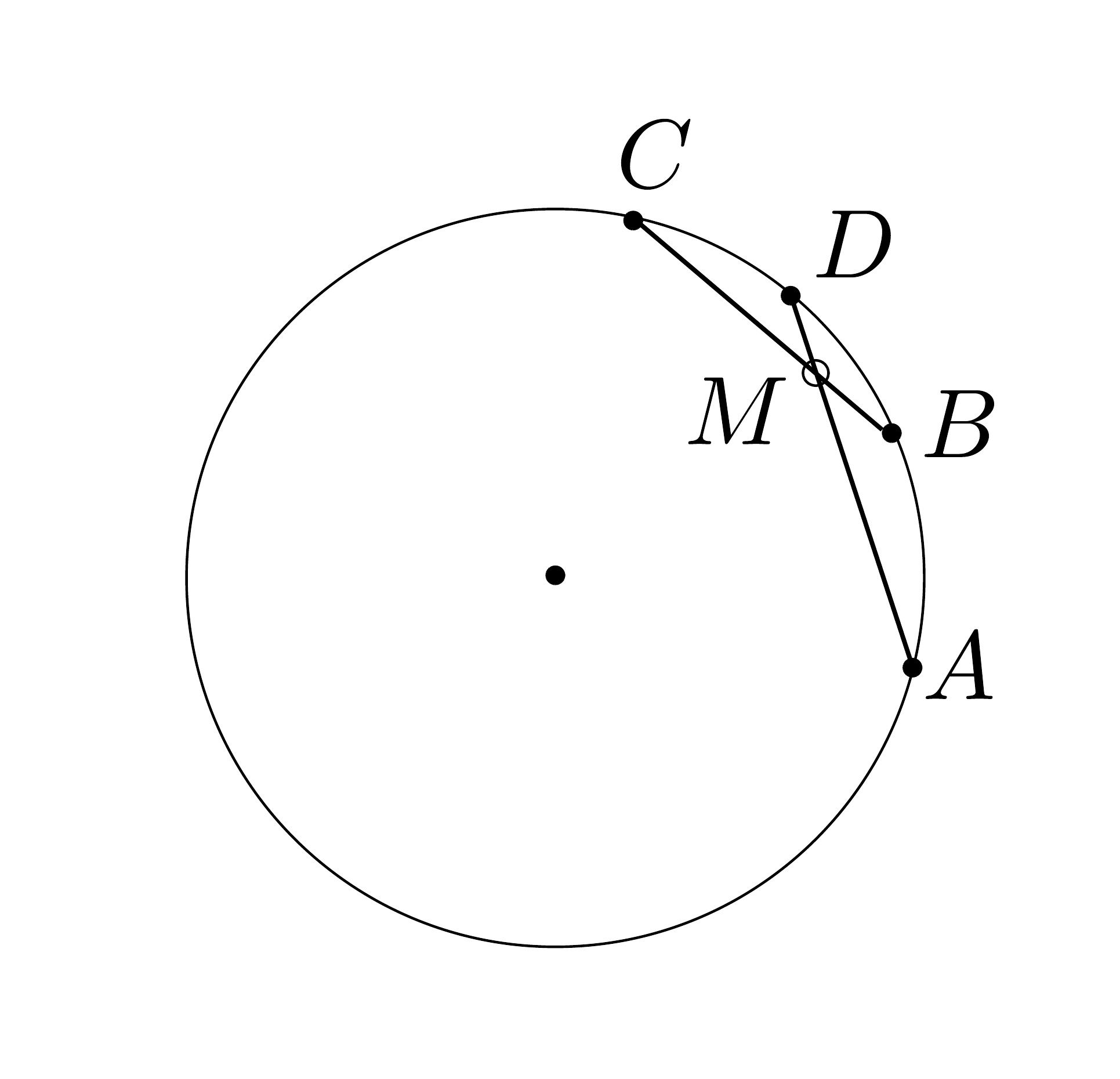}
\end{center}
\caption{{\bf Left:} The relative angles of the shown points are the same for each of the  three
circles. They represent the laser reflections $R(n_i)=(r_i,m_i)$ by their projected directions 
$r_i^\perp$ (norm $\sqrt{1-\rho^2}$), by their throat points $q_i$ (throat radius $\sigma_0$),
and by their projected moments $m_i^\perp$ (norm $\sigma_0 \rho$).
{\bf Right:} Four points with the same relative angles as in the left diagram, prepared for 
Theorem~\ref{th_affcomb}. If $M=tB + (1-t)C$ then $t$ parametrizes the affine combination
of $(A,B,C)$ that yields $D$.}
\label{fig_circleconfig}
\end{figure}

Proposition~\ref{prop_mom} immediately implies (the left of Figure~\ref{fig_circleconfig}):

\begin{corollary}
\label{cor_mom} 
If the origin is chosen to be the point on the mirror axis $A$ that is closest to the skewly incoming laser beam $L$, and if we denote the Pl\"ucker coordinates of two laser reflections by
$\pi(R(n_1))=(r_1,m_1)$ and $\pi(R(n_2))=(r_2,m_2)$ then
\begin{eqnarray*}
r_1\cdot r_A &=& r_2\cdot r_A (=\rho),\\
m_1\cdot r_A &=& m_2\cdot r_A (=\mu),\\
\left<{m_1^\perp,m_2^\perp}\right> &=& \left<{r_1^\perp,r_2^\perp}\right>,
\end{eqnarray*}
as oriented angles (viewed from $r_A$).
\end{corollary}

\section{Affine combination of cocircular points}
\label{sec_affcirc}

Using the assumptions and notations of Section~\ref{sec_Pl}, we have shown that for different
mirror positions ${\cal M}(n_1), {\cal M}(n_2), {\cal M}(n_3),\ldots$ we can consider three circles in the plane ${\cal D}_0$, centered at the origin (the left of Figure~\ref{fig_circleconfig}):
\begin{itemize}
\item containing the points $q_1, q_2, q_3, \ldots$
\item containing the direction projections $r_1^\perp, r_2^\perp, r_3^\perp,\ldots$ 
\item containing the moment projections $m_1^\perp, m_2^\perp, m_3^\perp,\ldots$
\end{itemize}
Furthermore, on each circle we observe identical oriented angles between points that correspond
to the same laser reflections $R(n_i)$ and $R(n_j)$, which is determined by the (rotation) angle
between $n_i$ and $n_j$ (by factor 2). As we will see, this implies that we can use the same
{\em affine combinations\/} for all these circles. In the next section we will prove that these
affine combinations on the circle can moreover be copy pasted to the Pl\"ucker coordinates of
the reflected lines.

Let $A, B, C$ be three non-collinear points in some plane, then we can uniquely express each
point $D$ in the (this) plane as an affine combination of $A, B, C$:
$$D = xA+yB+zC,\;\;\mbox{ with }x+y+z=1.$$
Because $z=1-x-y$ we count 2 dof for these combinations, which meets the number of dimensions
of the plane. Notice that $A, B, C$ determine a circumscribing circle ${\cal C}$. Now we will restrict ourselves in generating only points $D$ on this
circle ${\cal C}$, leaving us with only 1 dof for the coefficients $(x,y,z)$. In this section we will
express these coefficients as rational functions in a parameter that is explicitly 
determined by the relative angles between $A, B, C, D$.

The fundamental idea leading to our formulas is to parametrize the affine coefficients by
the location of the point of intersection $M$ of the lines $AD$ and $BC$ (the right of Figure~\ref{fig_circleconfig}). 

\begin{theorem}
\label{th_affcomb}
Let $a=|BC|$, $b=|AC|$ and $c=|AB|$ denote the edges of the triangle $ABC$, and let
$D=xA+yB+zC$ be a point on the circumscribing circle ${\cal C}$ of this triangle, with $x+y+z=1$.
If $M=AD\cap BC = tB + (1-t)C$ then
\begin{equation}
\label{eq_baryint}
(x\;\; y\;\; z) = \frac{(1\;\; t\;\; t^2)\cdot T}{(1\;\; t\;\; t^2)\cdot N},
\end{equation}
where
\begin{equation}
\label{eq_defTN}
T=\left(
\begin{array}{ccc}
0&0&-b^2\\ a^2&-b^2&2b^2-c^2\\-a^2&b^2-c^2&c^2-b^2
\end{array}\right)\;\;\mbox{ and }\;\;
N=\left(
\begin{array}{c}
-b^2\\ a^2+b^2-c^2\\-a^2
\end{array}\right).
\end{equation}
\end{theorem}

\noindent {\bf Proof:}
It can be proven that the necessary and sufficient condition on the barycentric
coordinates $(x,y,z)$ for $D$ to lie on the circumcircle ${\cal C}$ is given by (Fact 4 in \cite{Volenec}):
\begin{equation}
\label{eq_barycirc}
a^2yz + b^2 xz + c^2 xy = 0.
\end{equation}

Because $M=AD\cap BC$, this point can be given barycentric coordinates w.r.t. $\{A,D\}$ as well
as $\{B,C\}$ (Figure~\ref{fig_circleconfig}):
$$M=tB+(1-t)C=sA+(1-s)D.$$
Eliminating $M$ and solving for $D$ we obtain:
\begin{equation}
\label{eq_xyzst}
D = \frac{-s}{1-s}A+\frac{t}{1-s}B+\frac{1-t}{1-s}C.
\end{equation}
Since the sum of these coefficients equals 1, they must be equal to the barycentric coordinates $(x,y,z)$, expressed in function of $t$ and $s$. Substituting the barycentric coordinates of $D$
as given by Eqn.~\ref{eq_xyzst} in the circle condition of Eqn.~\ref{eq_barycirc},
we can solve for $s$:
\begin{equation}
\label{eq_sols}
s = \frac{a^2(t^2-t)}{(b^2-c^2)t - b^2}.
\end{equation}
Finally, substituting this expression for $s$ in Eqn.~\ref{eq_xyzst} yields the aimed claimed
formula in Eqn.~\ref{eq_barycirc}. 

\hfill $\blacksquare$

Observe that we do not lose generality by assuming that ${\cal C}$ equals the unit circle. Indeed, the affine coefficients remain invariant under scaling and
translations:
$$D=xA+yB+zC\Rightarrow wD+Z=x(wA+Z)+y(wB+Z)+z(wC+Z).$$
Furthermore, it can be easily seen that this affine combination is also not affected by rotations, such
that we can choose $A=(1,0)$. Consequently, the computation of the
coefficients $(x,y,z)$ in Eqn.~\ref{eq_baryint} only depends on the relative angles between the points.
\section{Affine combination of reflected lines of a single laser by a rotating mirror}
\label{sec_affpl}

Consider four laser reflections $R(n_i)$ by four mirror positions ${\cal M}(n_i)$ ($i=1,\ldots,4$).
From Theorem~\ref{th_hyperb} in Section~\ref{sec_refl}
we know that the lines $R(n_i)$ belong to a ruled surface of revolution,
a one-sheeted hyperboloid in general, or one of its degenerations in singular cases.
If $\pi(R(n_i))=(r_i,m_i)$ denote the Pl\"ucker coordinates, 
and if $r_i^\perp$ denote the projections of $r_i$ on ${\cal D}_0$,
then the relative angle of revolution between $R(n_i)$ and $R(n_j)$ can be written as 
\begin{equation}
    \theta_{ij}=\left<{r_i^\perp, r_j^\perp}\right>= 2\left<{n_i, n_j}\right>.\end{equation}

\begin{theorem}
\label{th_lines}
Let us represent the rotation angles of four laser reflections by points $P_1,\ldots,P_4$ on a
(unit) circle, that is, the arc between $P_i$ and $P_j$ equals $\theta_{ij}$. Then, the
affine combination $P_4=x_1P_1+x_2P_2+x_3P_3$ (with $x_1+x_2+x_3=1$) also applies to the
Pl\"ucker coordinates of the reflected lines:
\begin{equation}\pi(R(n_4))=x_1\pi(R(n_1))+x_2\pi(R(n_2))+x_3\pi(R(n_3)).\end{equation}
\end{theorem}

\noindent {\bf Proof.}
Let us first assume the origin at the centre $p$ of the ${\cal H}(L,A)$, which is a hyperboloid in general,
or a cone in case $L$ intersects $A$. For now, we exclude the degenerate case where $L$ intersects
 $A$ perpendicularly, implying that all reflections belong to the same plane.  
In Corollary~\ref{cor_mom} it is stated that the relative angles of the projected moments of the reflected lines $R(n_i))$ are identical to the relative angles of revolution (Section~\ref{sec_Pl}):
\begin{equation}\left<{m_i^\perp, m_j^\perp}\right>=\theta_{ij}=\left<{r_i^\perp, r_j^\perp}\right>.\end{equation}

So, 
\begin{eqnarray*}
r_4^\perp &=& x r_1^\perp + y r_2^\perp + z r_3^\perp\\
m_4^\perp &=& x m_1^\perp + y m_2^\perp + z m_3^\perp\\
\end{eqnarray*}
Furthermore, $\pi(R(n_i)) = (r_i^\perp + \rho r_A, m_i^\perp + \mu r_A)$. Using
$x+y+z=1$:
\begin{eqnarray*}
x \pi(R(n_1)) + y (\pi(R(n_2)) +z \pi(R(n_3)) &=&\\
(x  r_1^\perp + y r_2^\perp + z r_3^\perp +(x+y+z)\rho r_A,  x m_1^\perp + y m_2^\perp + z m_3^\perp + (x+y+z)\mu r_A) &=&\\
(r_4^\perp + \rho r_A, m^\perp_4 + \mu r_A) &=&\\
 \pi(R(n_4)) &&
\end{eqnarray*}
In case $L$ intersects $A$ perpendicularly, things become more simple. Then, the reflected lines
belong to a flat pencil, all assumed to intersect in the origin. In this case,
$\pi(R(n_i))=(r_i^\perp + \rho r_A,0,0,0)$, and hence the previous argument still holds,
restricted to the first three Pl\"ucker coordinates. 

Next, we drop the assumption about the location of the origin in 3-space. The general situation
can be transformed to the special situation (as described above) by a translation, which is
a linear transformation $T_4$ of $\Pbar^3$ (represented by a $4\times 4$ matrix). 
One can prove that this induces a linear transformation $T_6$ for the line coordinates $\pi(L)$
(represented by a $6\times 6$ matrix) \cite{PW}. The proof now is finished by the
fact that linear transformations preserve affine combinations.   

\hfill $\blacksquare$

\section{Data-driven calibration of rotating laser reflections}
\label{sec_algo}

The previous explanation enables to predict a laser reflection by a rotating mirror ${\cal M}(n))$, 
once three line reflections are known for three mirror positions. Notice that we bypass the geometry
of the incoming laser beam $L$ relative to the mirror axis $A$, neither do we need the spatial
position of the mirror plane that corresponds to an (initial) angle.
Notice that the described procedure equally well applies to devices with rotating
lasers instead of rotating mirrors.
\\
\\
{\bf input:} relative angles $\left<{n_i,n_j}\right>$ for three mirror positions 
${\cal M}(n_1), {\cal M}(n_2), {\cal M}(n_3)$, and the coordinates of the corresponding laser reflections:
$\pi(R(n_1)), \pi(R(n_2)), \pi(R(n_3))$.
\\
{\bf query:} $n_4$, or rather $\left<{n_i,n_4}\right>$ for some $i=1, 2, 3$.
\\
{\bf output:} $\pi(R(n_4))$.

\bigskip
\noindent
{\bf The algorithm:} 
\begin{enumerate}
\item Transform the mirror positions to rotation angles of the reflected lines:
$$\theta_{ij}=\left<{r_i^\perp, r_j^\perp}\right>=2 \left<{n_i,n_j}\right>.$$
\item Compute $T$ and $N$ as stated in Theorem~\ref{th_affcomb}
(Eqn.~\ref{eq_defTN}). This can be done by representing the three base angles and the fourth query angle on a (unit) circle, or directly
in terms of $\cos(\theta_{ij})$ and $\sin(\theta_{ij}))$.
\item Compute parameter $t$. Combine $t$, $T$ and $N$ to obtain the affine
coefficients $x, y, z$ (Eqn.~\ref{eq_baryint}).
\item Return  $\pi(R(n_4))=x\pi(R(n_1))+y\pi(R(n_2))+z\pi(R(n_3))$.
\end{enumerate}

\bigskip
\noindent
{\bf Algorithmic details:}
\begin{itemize}
\item When this algorithm is applied in a real world situation, we assume only small deviations from the mathematical conditions: the laser beam is always kept fixed, the rotation axis
for the mirror is always kept fixed, the mirror shape is close to a plane, the mirror reflection is close to perfect (hardly damaged by scratches and holes).
\item The first step of the algorithm may be more involved in certain practical situations. Indeed,
the control of the rotating mirror is done by user parameters $\omega_i$ that are not
necessarily equal to the geometric angles between the $n_i$. For instance, the mirror rotation might be galvanic driven, requiring input control in volts. The transformation from voltages to
geometric angles might or might not be linear. Even if the user is allowed to use angular values
for the input parameters, they are not necessarily identical to the geometric angles due to 
system noise. In this case, we obtain the angles directly from the measured reflection lines:  $\theta_{ij}=\left<{r_i^\perp, r_j^\perp}\right>$. The transformation
$\omega_{ij}=\omega_i - \omega_j \mapsto \theta_{ij}$ can be obtained by analytic or
probabilistic interpolation.
\item 
For the computation of the parameter $t$ it is recommended to permute $\{A,B,C\}$
in Theorem~\ref{th_affcomb} if needed, such that the chords $AD$ and $BC$ intersect inside the circle:
$M=AD\cap BC = tB + (1-t)C$. This guarantees that $t\in [0,1]$ and significantly improves
the stability.\\
\end{itemize}

\section{The 3 by 3 line grid calibration of a 2M-GLS}
\label{sec_vibro}

This section is motivated by a {\em galvanometer\/}, a sensor consisting of a single fixed laser $L$ that is internally reflected by two sequential mirrors, 
each rotating about an individual axis, denoted by $A$ and $B$ in order of reflection. The control
of the individual rotating mirrors is typically galvano-driven, allowing two independent user parameters, denoted by
$\alpha$ and $\beta$ respectively (Figure~\ref{fig_vibro}).
As explained in Section~\ref{sec_algo}, we may assume that we can express mirror angles in radians.

Assume for the moment that we fix the second mirror at angle $\beta_1$.
By means of the algorithm of Section~\ref{sec_algo}, we can predict an outgoing line $R(\alpha,\beta_1)$ by means of three observed lines $R(\alpha_1,\beta_1)$, 
$R(\alpha_2,\beta_1)$ and $R(\alpha_3,\beta_1)$ corresponding to three positions of the
first rotating mirror ${\cal M}(A,\alpha)$:
\begin{equation}\pi(R(\alpha,\beta_1)) = 
x_\alpha \pi(R(\alpha_1,\beta_1)) + y_\alpha \pi(R(\alpha_2,\beta_1)) + 
z_\alpha \pi(R(\alpha_3,\beta_1)),\end{equation}
where the affine coefficients $(x_\alpha, y_\alpha, z_\alpha)$ are computed
by Formula~\ref{eq_baryint}.
Note that these coefficients do not depend on the specific choice $\beta_1$ for the position of the
second mirror. Indeed, the relative angle $\theta_{ij}$ between $(R(\alpha_i,\beta_1)$
and $(R(\alpha_j,\beta_1)$
is the opposite of the corresponding relative angle on ${\cal H}(L,A)$. More precisely,
$$|\theta_{ij}| = 2|\alpha_j - \alpha_i|.$$

%

\begin{theorem}
A two-mirror galvanometric laser scanner is intrinsically calibrated by the knowledge of $3\times 3$ emitted lasers
$R(\alpha_i,\beta_j)$ corresponding to a grid of $3\times 3$ combinations of mirror pairs
$(\alpha_i,\beta_j)$ ($i=1,2,3$ and $j=1,2,3$).
\label{th_calib2M}
\end{theorem}

\noindent {\bf Proof.}
We show that for each given query pair $(\alpha, \beta)$, we can predict the corresponding
double reflected laser $R(\alpha, \beta)$ by means of the given laser grid. To this end, we first
compute the affine coefficients $(x_\alpha, y_\alpha, z_\alpha)$ for a fixed $\beta_j$.
In principle, the resulting coefficients are identical for each choice of $\beta_j$ ($j=1,2,3$).
Consequently, we obtain:
\begin{eqnarray*}
\pi(R(\alpha,\beta_1)) &=& 
x_\alpha \pi(R(\alpha_1,\beta_1)) + y_\alpha \pi(R(\alpha_2,\beta_1)) + 
z_\alpha \pi(R(\alpha_3,\beta_1)).\\
\pi(R(\alpha,\beta_2)) &=& 
x_\alpha \pi(R(\alpha_1,\beta_2)) + y_\alpha \pi(R(\alpha_2,\beta_2)) + 
z_\alpha \pi(R(\alpha_3,\beta_2)).\\
\pi(R(\alpha,\beta_3)) &=& 
x_\alpha \pi(R(\alpha_1,\beta_3)) + y_\alpha \pi(R(\alpha_2,\beta_3)) + 
z_\alpha \pi(R(\alpha_3,\beta_3)).
\end{eqnarray*}
These three laser lines belong to a system of rulers of the hyperboloid ${\cal H}(L(\alpha),B)$, defined
by the mirror axis $B$ and the incoming laser beam $L(\alpha)$, which is the reflection of $L$
by the $\alpha$-position of the first mirror. Applying the algorithm van Section~\ref{sec_algo}
once more, we obtain:
$$\pi(R(\alpha,\beta)) = 
x_\beta \pi(R(\alpha,\beta_1)) + y_\beta \pi(R(\alpha,\beta_2)) + 
z_\beta \pi(R(\alpha,\beta_3)) .$$

\hfill $\blacksquare$

\section{Experiments}
\label{sec_exp}

\begin{figure}[htbp]
\includegraphics[width=8cm, height=8cm]{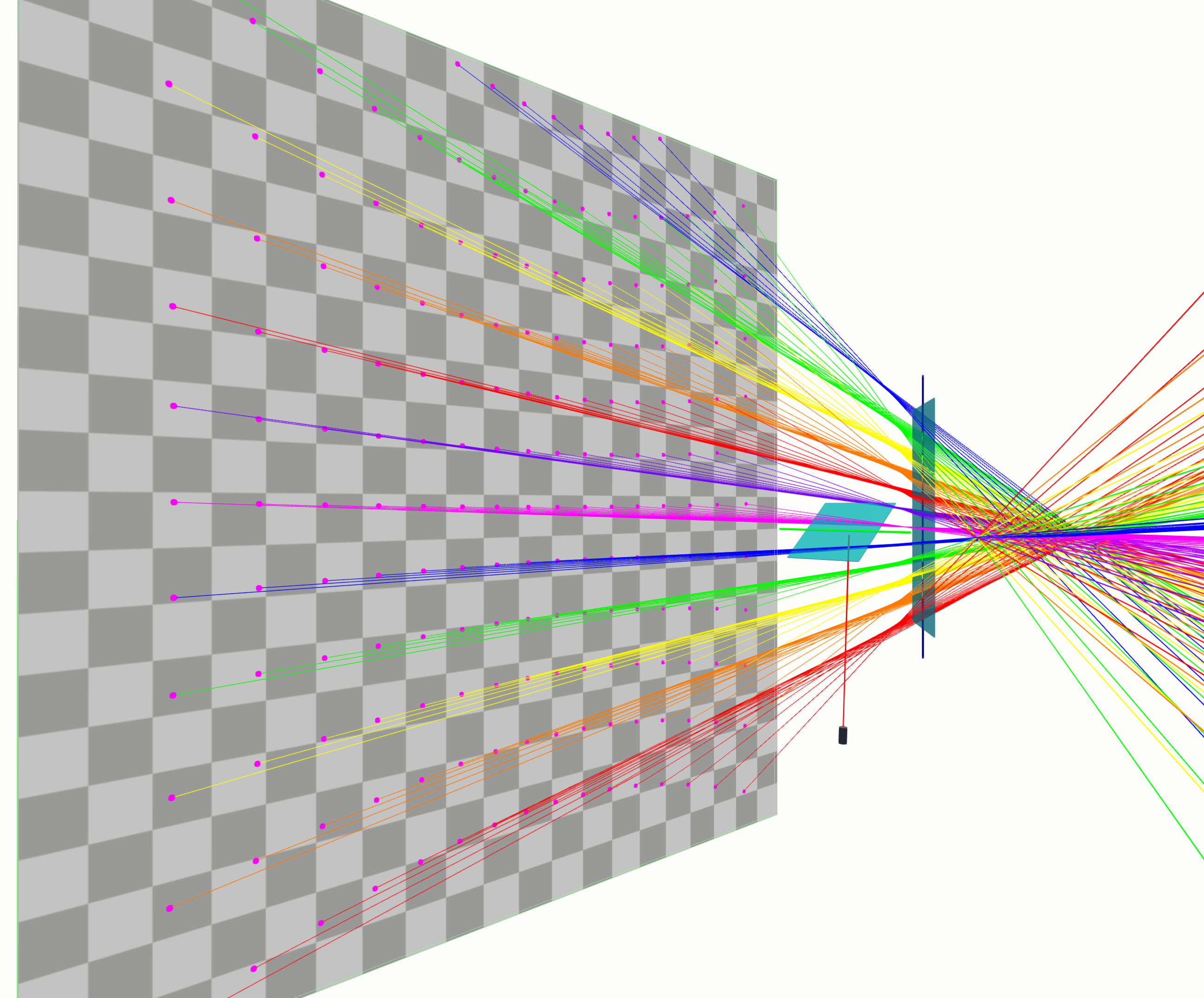}
\centering
\caption{The virtual setup. A laser beam is reflected by two rotating mirrors. The reflected laser beams hit a detection plane. The 3D coordinates of the points (the pink dots) are recorded.}
\label{fig:gen_data}
\end{figure}

In order to validate our hyperboloid grid model, we apply the method of Section \ref{sec_vibro} to synthetically generated data. The aim is to predict the set of Pl\"ucker coordinates for a given pair of mirror rotations. The benefit of working with synthetic data is that an exact underlying ground truth can be established. To this end, we built a setup in a virtual environment in the game engine Unity (version 2020.2.5f1). We placed two rotating mirrors and a laser in a configuration that can also be found in for instance a Polytec PSV-400 laser Doppler vibrometer (Figure~\ref{fig:gen_data}). 
	A real time demonstration of the setup in which the mirrors rotate to reflect an incoming laser beam can be seen in 
 \url{https://youtu.be/GNTjmJvdTpw}.
We generated laser beams for twelve rotation angles for the first mirror and sixteen for the second mirror, resulting in a 
$12\times 16$ grid of 192 lines.

\begin{figure}[htbp]
\includegraphics[width=8cm, height=8cm]{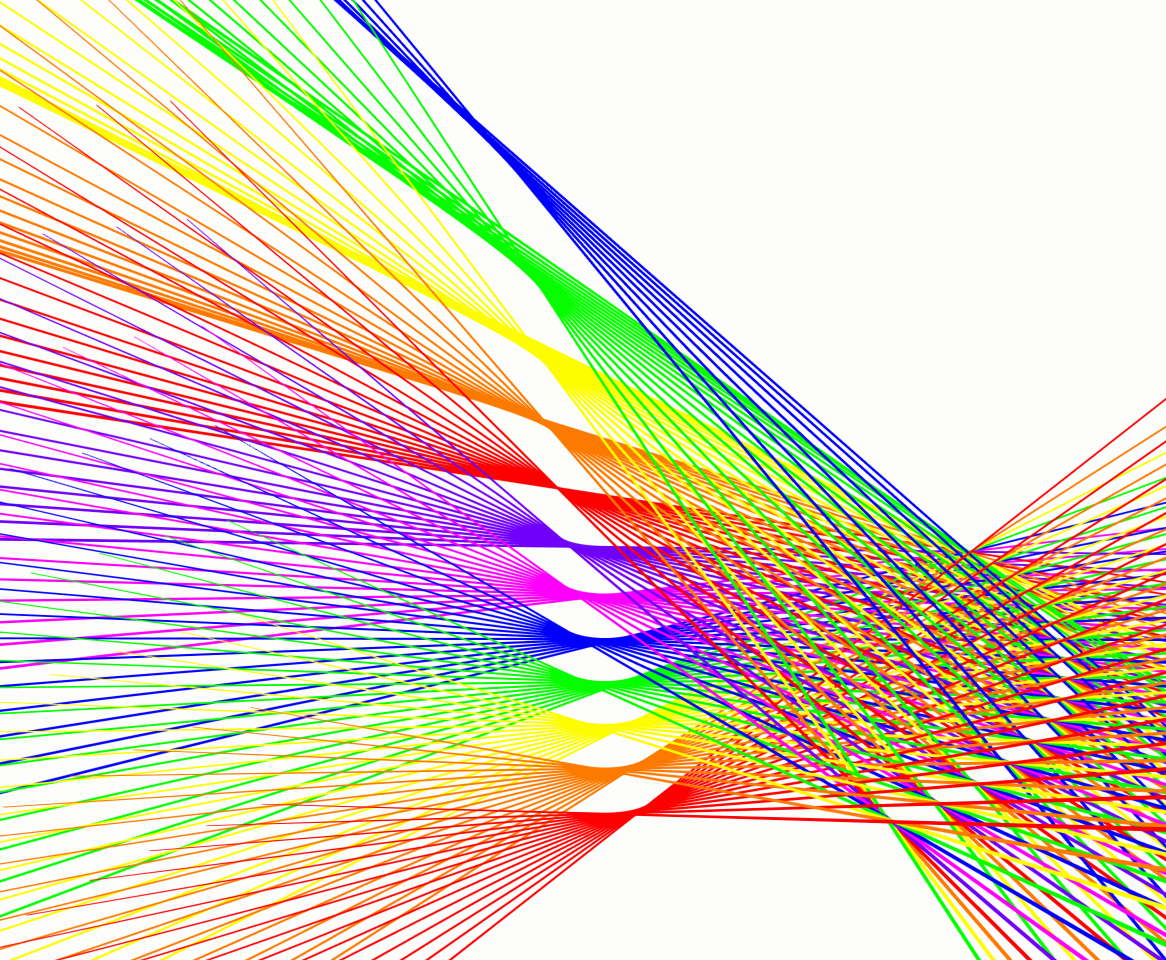}
\centering
\caption{Lines rotated around a central axis form a hyperboloid. In a galvanometric setup, the first mirror rotation defines which hyperboloid, while the second mirror rotation determines the line on that hyperboloid.}
\label{fig:hyp}
\end{figure}

To measure the Pl\"ucker coordinates of those (reflected) laser beams, we placed a detection plane in front of the setup and recorded where the laser beams intersect that plane. All reflections and the detection of intersections are handled by the built-in Unity physics engine. An overview of the virtual setup can be found in Figure~\ref{fig:gen_data}. The reflected laser beams for a set of co-axial hyperboloids are visualised in detail in Figure~\ref{fig:hyp}. The detection plane is then moved and rotated in eight positions. The simulation scale is chosen such that the distances of the
detection planes vary from approximately 1000 to 2600 millimetre. 

Consequently, for each pair of mirror rotation angles (which uniquely generate a single laser beam), we obtain eight points.
Strictly speaking, we only need to put the detection plane in two positions. However, to simulate real world conditions, we added Gaussian noise to the 3D coordinates of the detected points. We performed our simulations at seven noise levels with standard deviations equal
to 0, 1, 2, 4, 6, 8 or 10 millimetres. For each of the noise levels, we generated 50 sets of lines.
We average our findings over those 50 sets to eliminate statistical artefacts in the noise of the data.
We perform a best fit method as described in \cite{LSGC} to calculate the Pl\"ucker coordinates for the straight line generated by the mirror rotation pairs.

For each noise level we select a {\em basegrid\/},
being a subgrid of lines from the $12\times 16$ dataset.
We consider the following basegrid sizes: 
$3\times 3$, $4\times 4$, $6\times 6$, $8\times 8$ and $8\times 11$. 
This allows us to investigate the influence of the number of training lines on the accuracy of the calibration. To avoid unnecessary numerical
problems, the angles in these subgrids are (uniformly) spread out in the range of the
rotation angles of the sensor mirrors.
The angle pairs of the data sets that do not participate in the basegrid provide a test set, for which we use the Unity-generated lasers with zero noise as ground truth. The aim now is to predict the lines in the test sets when only the two mirror rotation angles are given.  The proposed method uses the base grid to recover the two-mirror congruence as a double system of hyperboloids of revolution (Corollary~\ref{cor_hypernet}). 
This line congruence is compactly represented as a $3\times 3$ grid
that enables laser predictions by means of affine grid combinations (Theorem~\ref{th_calib2M}).

\begin{figure}[htbp]
\includegraphics[width=9cm, height=9cm]{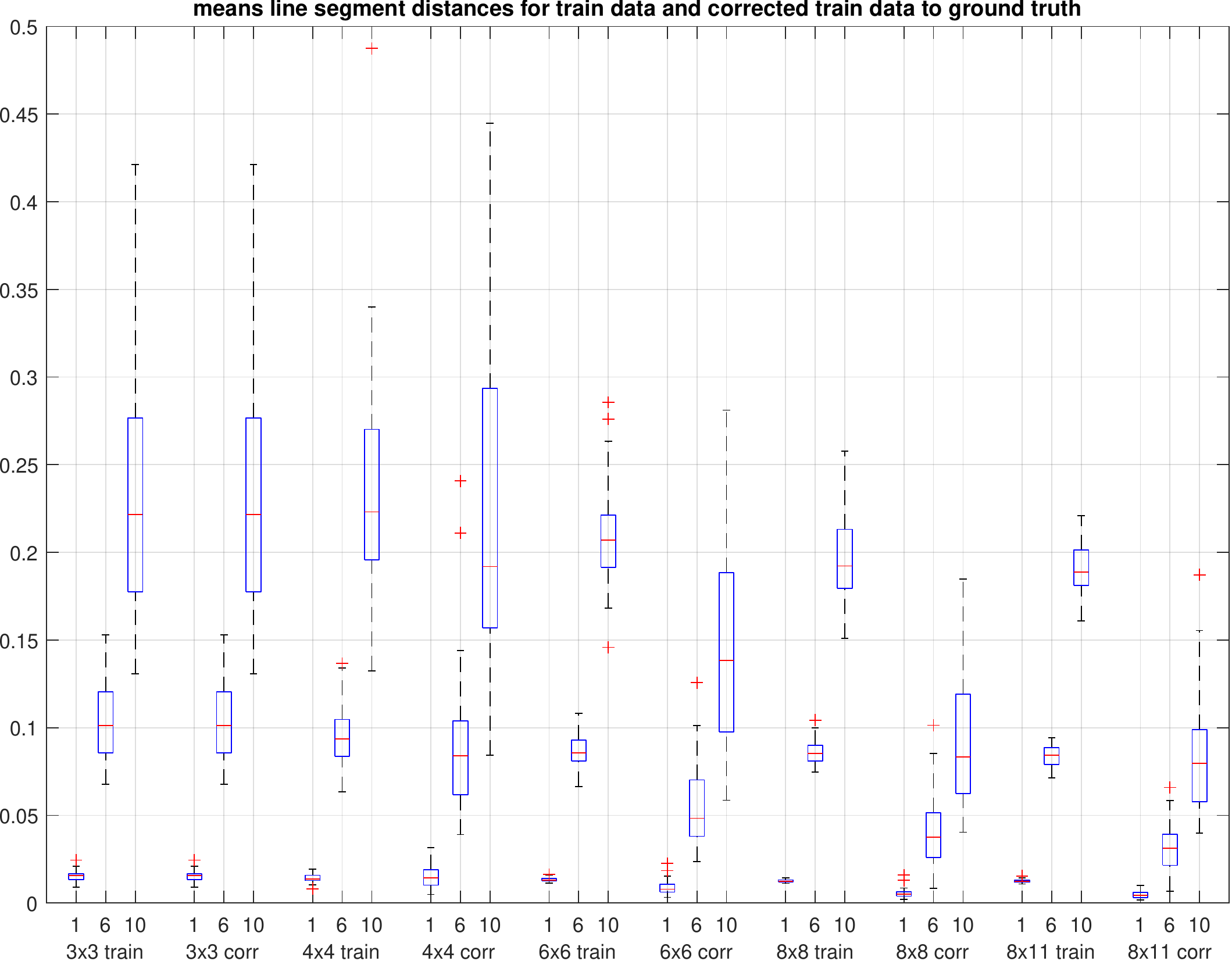}
\centering
\caption{The means of the line segment errors with respect to the ground truth, for the measured lines as well as for the corrected lines (by
hyperboloid grid fitting). The boxplots are grouped by five grid sizes and within each group ordered by three noise levels during the measurement of the 8 points (at a distance
of at most 3 m) that are used for the line measurements: standard
deviations of 1, 6 and 10 mm.}
\label{fig_hypgridcorr}
\end{figure}

A procedure for a robust fitting of a hyperboloid grid to a basegrid of noisy line measurements is described in Appendix~\ref{app_hypcor}.
Because the quality of this fitting has a significant share in the accuracy of our method, we present it here as an intermediate validation
in the framework of the previously described synthetic experiment. The results are shown in 
Figure~\ref{fig_hypgridcorr}.
The gain (noise reduction) is most apparent for grids ranging $6\times 6$ and up.

We compare our method to the semi-data driven method described in \cite{IvanConstrGP}, 
where the authors validate the performance and feasibility of semi-data driven approaches by means of Gaussian processes. The method of \cite{IvanConstrGP} outperforms current state-of-the-art physical-based calibrations, and performs at least equally well as other existing statistical or machine learning methods, which makes it an appropriate reference to compare our method with.

Following the procedure in Section D
of \cite{IvanConstrGP}, a Gaussian process  is trained for each of the six components of the Pl\"ucker coordinates \cite{GPinLearning}. In the implementation of the Gaussian processes, we used
a periodic kernel with automatic relevance determination as suggested by \cite{IvanConstrGP}:
\begin{equation}
	\begin{split}	 	
		k_{PER}(\mathbf{x}, \mathbf{x}') = \sigma^2_{f}
		&\exp \left(- \frac{2}{l^2_\alpha} \sin^2\left(\frac{\left|\alpha - \alpha'\right|}{2} \right) \right) 
		\\& \cdot \exp \left(- \frac{2}{l^2_\beta} \sin^2\left(\frac{\left|\beta - \beta'\right|}{2} \right) \right) .
		\label{PER2}
	\end{split}
\end{equation}
 
\begin{figure}[htbp]
\includegraphics[width=9cm, height=9cm]{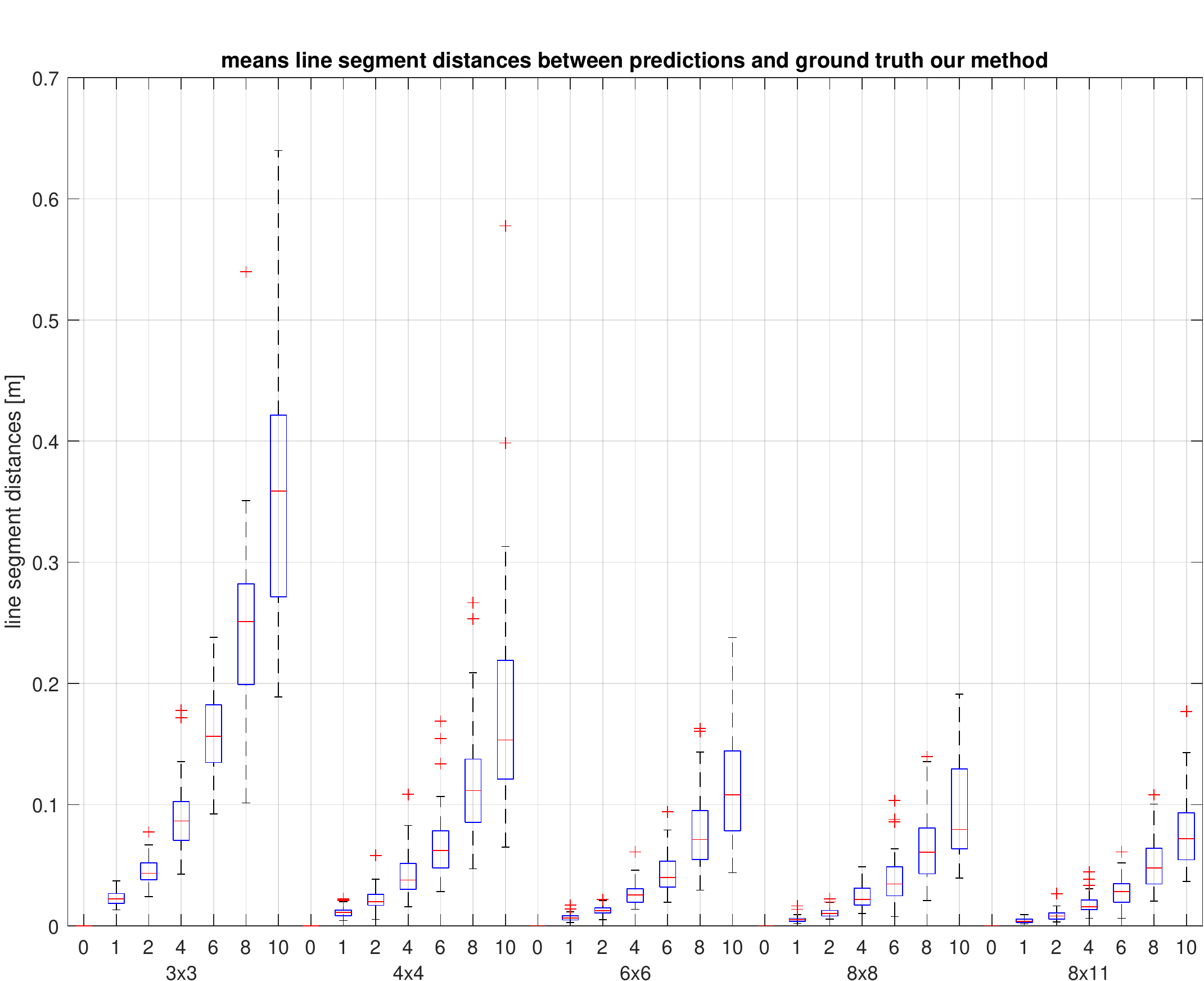}
\centering
\caption{The means of the line segment distances between the predicted test lines (by the
proposed method) and the ground truth. The boxplots are grouped by five grid sizes and within each group ordered by seven noise levels during the measurement of the 8 points
(at a distance of at most 3 m) that are used for line fitting: standard
deviations of 0, 1, 2, 4, 6, 8 and 10 mm.}
\label{fig:anal_alg}
\end{figure}

\begin{figure}[htbp]
\includegraphics[width=9cm, height=9cm]{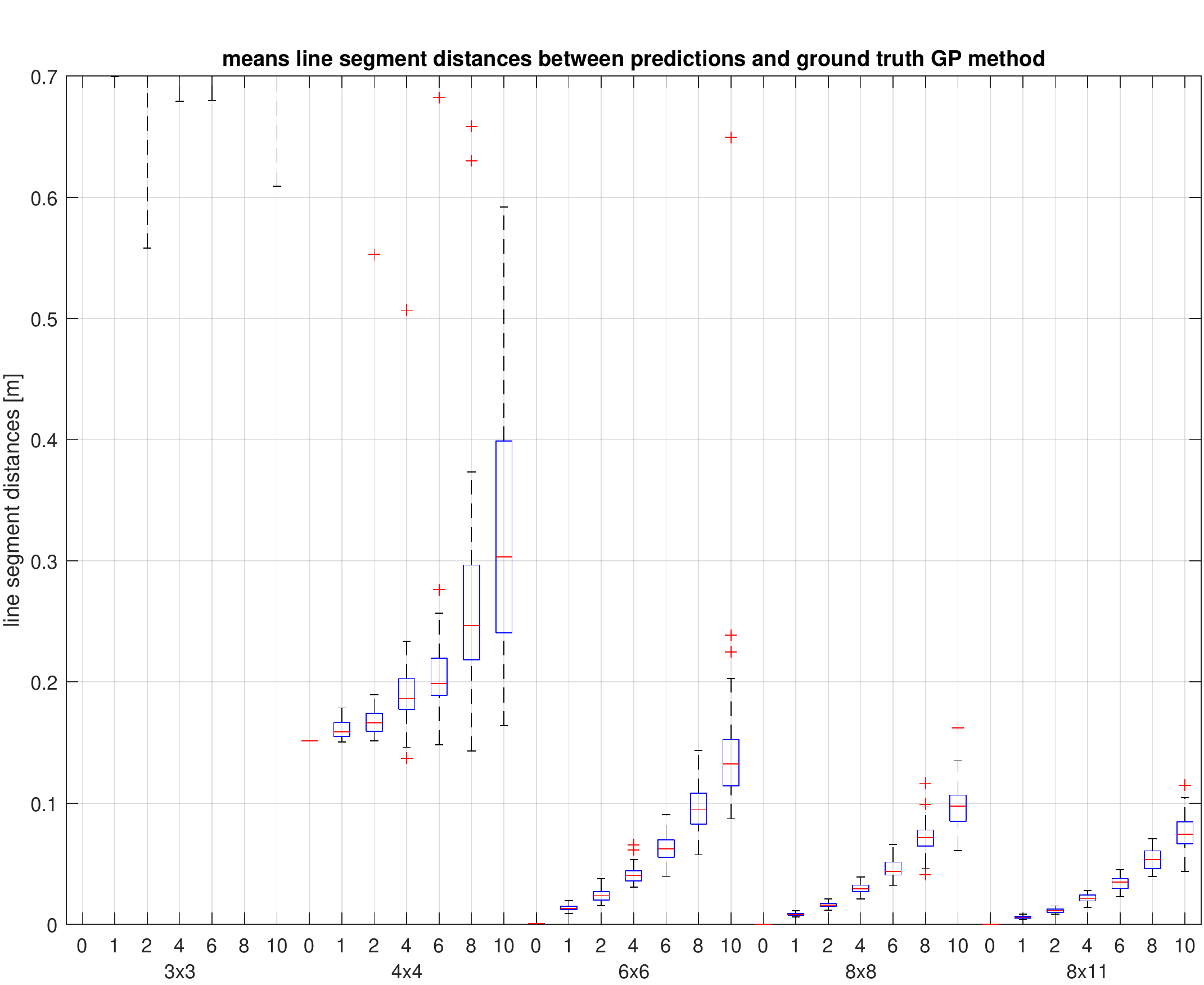}
\centering
\caption{The means of the line segment distances between the predicted test lines (using a GP)
and the ground truth. The boxplots are grouped by five grid sizes and within each group ordered by the seven noise levels. For the minimal training set ($3\times 3$), the 
GP-model predicts values so far from the ground truth that they are no longer of any significance. The data has become to sparse to work with.}
\label{fig:anal_GP}
\end{figure}

In order to evaluate the prediction quality of any method, we need a measure for
the difference between two spatial lines. In our experiments we worked with several distance measures, but they appeared to agree with respect to the final conclusions. In the
presentation of our results, we use the line distance measure as suggested by \cite{PW}. 
For the computation of this measure, we need to define two fixed parallel planes,
with the certitude to limit our region of interest. 
As a matter of fact, we chose them perpendicular to the Z-axis (more or less
the direction of the outgoing beams), 
one through the origin, the other at a distance of 10 metres. Two lines intersect these two planes in four points $\mathbf{g_1}, \mathbf{g_2}, \mathbf{h_1}$ and 
$\mathbf{h_2}$  (same indices for the same line, some letters for the same plane).
We calculate the so called {\em line segment distance\/} $d$ as follows:
\begin{equation}	 	
	d^2 =  ||\mathbf{g_1}-\mathbf{g_2}||^2 + ||\mathbf{h_1}-\mathbf{h_2}||^2 + (\mathbf{g_1}-\mathbf{g_2})\cdot(\mathbf{h_1}-\mathbf{h_2}) .
\end{equation}
The line segment distance is computed to evaluate the error of each predicted line with respect
to its ground truth. We took the average for all the lines per test set. This results in 50 averages per combination of grid size and noise level. This is done for the proposed method as well as for the GP-method. An overview of the results can be found in Figure~\ref{fig:anal_alg} and in Figure~\ref{fig:anal_GP}. Note that the line segments that we used in our error measure have a length of at least 10 metre, which should be taken into account in the interpretation of the prediction error on the vertical axis of the figures (expressed in metre) For example, an lsd-error of $0.1$ m
for a predicted line is a line segment deviation of at most 1 cm per metre. 

The runs with zero noise confirm that the proposed hyperboloid grid calibration is an exact method,
even when using a minimal $3\times 3$ grid. We also notice that for a measurement noise expressed by a standard deviation of $\sigma$ mm (within a work space of 2 to 3 m), 
the line prediction error appears to be lower than $2\sigma$ mm (per metre) assuming a basegrid size of
at least $4\times 4$, and even bounded by $\sigma$ mm (per metre) if we use basegrids of size $6\times 6$ or larger. From our experiments there seems to be no convincing motivation to use basegrid
sizes larger than $8\times 8$. On the other hand, we observe that boxes are stretched out (between first and third quartiles) in cases where point measurements suffer from large noise levels 
($\sigma > 7$ mm within the workspace region). This is explained by the fact that the basegrid data is corrected and fixed by a hyperboloid grid fit, such that the prediction errors for every test line are determined by the quality of this fit (Appendix~\ref{app_hypcor}),
which can be an unlucky estimate if the data noise happens to be unfortunate.

If we investigate the results of the GP-method for the same Unity-data (Figure~\ref{fig:anal_GP}), then we observe that $3\times 3$ grids are too small to teach a useful Gaussian process. 
Its performance takes over the proposed method from the moment the GP is trained by basegrids larger than $8\times 8$. In case of larger measurement noise, the variance of the GP results appears
to be smaller than for the proposed method. This is due to the fact that a Gaussian process keeps on balancing the measurement noise during the prediction of the test lines.

The datasets generated and analysed during the current study are publicly available in the github\\ repository \url{https://github.com/IvanDeBoi/Line-Calculus-on-Hyperboloids}.

\section{Conclusions and further research}
\label{sec_concl}

This paper offered a completely new method for the modeling and 
3D calibration of a galvanometric laser scanner with two mirrors.
As a matter of fact, 
the proposed calibration paradigm applies to any laser scanner with
rotational symmetry, such as other galvanometric systems or a Lidar,
sensors with a rapidly growing number of applications.
Our study provides a deeper understanding how
many sensors can be naturally represented as a specific line variety,
and how it pays off to discover the type of this variety by
a mathematical analysis. The proposed line model is more specific than
previously published general line models, but the calibration 
merely consists of measured line data, and does not need
to recover intrinsic parameters of a physical model.

As a main contribution we model a two-mirror-GLS as a 
{\em hyperboloid grid congruence\/} that can be represented in a compressed way by a $3\times 3$ basegrid of data lines. This is a significant simplification compared to the use of lookup tables commonly used in the 2D or 3D calibration
of a GLS.
We derived a formula that translates angular control parameters into simple affine combinations of these $3\times 3$ grid lines,
enabling our calibration model to make fast predictions.

In a follow-up article we describe how this formula allows us to find
an analytic solution for the reverse engineering problem: how to determine the pair of mirror angles that generate the laser reflection that hits a given 3D target point.

The hyperboloid grid model for a two-mirror galvanometric laser scanner and the affine combination formula for the $3\times 3$ grid is an interesting theoretical result,
but in order to validate its practical performance, and in order to compare it to a statistical training model (GP-method), we chose to fit a hyperboloid grid on larger training grids.
To this end, we designed a new algorithm for the robust fitting of a hyperboloid of revolution on given rulers. As it is the case for every regression model, this
choice implies the advantage of noise reduction, but the disadvantage of neglecting noise. Fitting on training grids of size at least $6\times 6$ causes line prediction errors
that are comparable or smaller than the point measure errors.

The results of the GP-method are inferior to the proposed methods for small grid sizes and for limited point measure noise levels. If the noise level is represented by a standard deviation of
8 mm or more (in the work space region up to 3 m), and if a basegrid is used of size at least $8\times 8$, the GP-method performs more accurately and more precisely.
This is due to the fact that a Gaussian process can be seen as a universal smoother, excellent at filtering out noise. On the other hand, the GP-method trains 6 separate line coordinates, and most
often they do not satisfy the Grassmann-Pl\"ucker relation. Consequently, it fails to deliver an effective line. This can be taken care of by post-corrections, or by using a GP with manifold constraints, but it is an additional complication. In \cite{IvanConstrGP} it is shown that the violation of the Grassmann-Pl\"ucker relation becomes less apparent when using larger training sets.

Also, if the mirror quality of a real galvanometric laser scanner significantly
deviates from our ideal mathematical assumptions, the GP-predictions will be more accurate than the idealised hyperboloid grid predictions.
On the other hand, discrepancies between the ideal predictions of our method and an observed laser from the real world sensor can detect defects or flaws in the device. This suggests that our line model 
can also be applied as a tool for quality control. 

\begin{appendices}
\section{Fitting a one-sheeted hyperboloid of revolution to given noisy rulers}
\label{app_hypcor} 

Examples of approximation methods for ruled surfaces are presented in \cite{RotHelixApprox,3DreconPott,approxsurfPott}.
However, in these approaches, the ruled surfaces are fitted to given point data, rather than to measured lines, as is the case in our situation.

Let ${\cal L}=\{L_1, L_2,\ldots, L_n\}$ be the noisy data lines
that are supposed to be rotated images of some (unknown) line around some (unknown) axis $A$. We assume that the $l_i$ are presented by their normalised Pl\"ucker coordinates:
$L_i = (r_i , m_i)$.  

{\bf Step 1:}\\
Considered as points, the correct normalised directions $r_i$ belong to a circle
centred at a point $x\in A$, in a plane ${\cal D}_x$ perpendicular to $A$. So, the direction
vector $r_A$ of $A$ can be recovered as the normal of a fitting plane. This plane ${\cal D}_x$
can be approximated by a robust technique such as
RANSAC or MLESAC \cite{RANSAC,MLESAC}. Observe that at this stage we only need to
recover the normal $r_A$ of ${\cal D}_x$.\\
In Section~\ref{sec_vibro} we have measurements of rulers of several hyperboloids 
${\cal H}(L(\alpha),B)$ at our disposal, all sharing the same axis $A$ of revolution, enabling us to
increase the accuracy of the direction of this axis by computing the median or a trimmed mean
of the computed $r_A$ of the individual hyperboloids ($||r_A||=1$).\\
It is an option to neglect from now on data lines $L_i$ of which the directions $r_i$
have been considered as outliers in the previous step. 

{\bf Step 2:}\\
Once we have found a reliable $r_A$, we can recover the pitch $\rho$ of the hyperboloid as the mean of the dot products $r_i\cdot r_A$. We ensure
that the directions are oriented such that all these dot products have positive signs. In this way,
we can reduce the noise on the axial component 
$r_i^\parallel = (r_i\cdot r_A)r_A$ of the line directions $r_i$: 
\begin{equation}r_i^\parallel \mapsto \rho r_A.\end{equation}

{\bf Step 3:}\\
We can also ``correct'' the rotational components $r_i^\perp=r_i - r_i^\parallel$, provided that we know the
relative angles $\theta_{ij}$ between each pair, what we normally do in case of a reliable
control of the involved galvanometric device. We proceed as follows. 
Based on the measured data, and the previously recovered axis direction $r_A$, we can
 consider the noisy projections $r_i^\perp = r_i - (r_i\cdot r_A) r_A$. The norms
$||r_i^\perp||$ will be all corrected as $\sqrt{1-\rho^2}$, so we can focus on their directions $d_i = r_i^\perp/||r_i^\perp||$, points on the unit circle in the plane ${\cal D}_o$ perpendicular
to $r_A$. We can represent them by a radian parameter $pr_i$.
The direction noise of the projections $r_i^\perp$ is reduced by solving a constrained optimization
problem in $n$ unknowns $cpr_i$, representing the corrected radian parameters $pr_i$.
More precisely, we minimise the sum of squared distances between $cpr_i$ and the 
noisy $pr_i$, constrained by the given relative angles $\theta_{ij}$. Actually, a closed form solution for
the $cpr_i$ is obtained by means of Lagrange multipliers. Finally, we map the $cpr_i$ back to
unit vectors $cd_i$ in the plane perpendicular to $r_A$, The
correction of the noisy directions $r_i$ of the data lines is done as follows:
\begin{equation}r_i \mapsto \sqrt{1-\rho^2} cd_i +  \rho r_A.\end{equation}

{\bf Step 4:}\\
Having a reliable direction $r_A$ at our disposal also facilitates the recovery of the complete
axis $A=(r_A,m_A)$. To this end, we lean on the property that the exact normalised rulers
$EL=(er,em)$ of the same regulus of a ruled surface of revolution
have a constant {\em bilinear product\/} with the
exact normalized axis $EA=(er_A, em_A)$, given by \cite{PW}
\begin{equation}\Omega(EL, EA) = er\cdot em_A + em\cdot er_A.\end{equation}
This motivates us to find the corrected $m_A$ as a Least-Squared Approximation for the
equations 
\begin{equation}0=\Omega(L_i,A)-\Omega(L_j,A)=(m_i-m_j)\cdot r_A + (r_i-r_j)\cdot m_A,\end{equation}
augmented with the Grassmann-Pl\"ucker relation for the line $A$: $r_A\cdot m_A = 0$, which might be multiplied by a weight factor if one needs to increase the importance 
to deliver a real line $A$. All these equations are linear, because we assume that
$r_A$ is known. 
In composing this overdetermined system of equations,
 we use the noisy data $(r_i,m_i)$ of the measured inliers $L_i$, and the recovered
direction $r_A$. Notice that we really need to obtain $r_A$ in a previous step, because
the equations $\Omega(L_i,A)-\Omega(L_j,A) = 0$ are not sufficient to determine $A$.
For example, rulers $R$ in the second regulus of the same hyperboloid satisfy
$\Omega(EL,R)=0$ for each ruler $EL$ of the first regulus.\\
Because a set of three rulers $\{ L_1,L_2,L_3 \}$ can serve for a minimal solver that recovers
the moment $m_A$ of $A$, we encounter here another opportunity to integrate a robust consensus procedure by random sampling, now eliminating rulers $L_i$ with moment ouliers. Furthermore, to avoid numerical instabilities, we recommend selecting the pairs
$\{ L_i,L_j \}$ for the equations $\Omega(L_i,A)-\Omega(L_j,A) = 0$ such that we
maximize $||r_i-r_j||$.

{\bf Step 5:}\\
Next, for each of the data lines $L_i$, we can compute its
(perpendicular) distance $\sigma_i$ to the recovered axis $A$, and its closest 
point $p_i$ on $A$. Without noise, all these points $p_i$ coincide with the centre $p$ of the
gorge circle of the hyperboloid, and all these distances $\sigma_i$ are equal to its radius $\sigma$.
So, we recover $p$ and $\sigma$ as the means of the $p_i$ and the $\sigma_i$ respectively.\\
In Section~\ref{sec_vibro} we have measurements of rulers of several hyperboloids 
${\cal H}(L(\alpha),B)$ at our disposal, all sharing the same axis, enabling us to apply
the previous steps for each of them, yielding a gorge centre on the recovered axes of these hyperboloids. The mean of these gorge centres provides a stable point $p_s$ on the common axis, that gives rise to a more accurate computation for the moment
as $m_A = p_s\times r_A$. \\
In any case, we use the reconstructed axis $A$ to approximate the centre $p$ and the radius
$\sigma$ of the gorge circle ${\cal C}_p$, enabling us to recover the gorge points of the
exact rulers $EL_i$: $q_i={\cal C}_p\cap EL_i$. Indeed, the directions $pq_i$ are exactly
the quarter turns of the corrected projected directions $cd_i$ (Proposition~\ref{prop_georefl}),
which together with the condition $||pq_i||=\sigma$ determines the location of $q_i$.

{\bf Step 6:}\\
Finally, we correct the moments $m_i$ of $L_i$ by $q_i\times r_i$, where we use the
corrected $r_i$.
\end{appendices}

\section*{Disclosures}
The authors declare no conflicts of interest.

\section*{Data availability}
A real time demonstration of the virtual experimental setup can be seen in\\
\url{https://youtu.be/GNTjmJvdTpw}.
The datasets generated and analysed during the current study are publicly available in the github repository\\ \url{https://github.com/IvanDeBoi/Line-Calculus-on-Hyperboloids}.

 \section*{Dedication}
 This article is dedicated to the memory of Henry Crapo, who introduced the first author to the fascinated world of projective line geometry.

 \bibstyle{alpabetic}
\bibliography{mybib}

\end{document}